# Non-sinusoidal current-phase relationship in Josephson junctions from the 3D topological insulator HgTe

Ilya Sochnikov[1,2], Luis Maier[4], Christopher A. Watson[1,3], John R. Kirtley[2,3], Charles Gould[4], Grigory Tkachov[5], Ewelina M. Hankiewicz[5], Christoph Brüne[4], Hartmut Buhmann[4], Laurens W. Molenkamp[4], and Kathryn A. Moler[1,2,3]

[1]*Department of Applied Physics, Stanford University, Stanford, California 94305, USA*

[2]*Geballe Laboratory for Advanced Materials, Stanford University, Stanford, California 94305, USA*

[3]*Stanford Institute for Materials and Energy Sciences, SLAC National Accelerator Laboratory, Menlo Park, California 94025, USA*

[4]*Physikalisches Institut (EP3), University of Würzburg, Am Hubland, 97074 Würzburg, Germany*

[5]*Institute for Theoretical Physics and Astrophysics, University of Würzburg, Am Hubland, 97074 Würzburg, Germany*

**We use Superconducting QUantum Interference Device (SQUID) microscopy to characterize the current-phase relation (CPR) of Josephson Junctions from 3-dimentional topological insulator HgTe (3D-HgTe). We find clear skewness in the CPRs of HgTe junctions ranging in length from 200 nm to 600 nm. The skewness indicates that the Josephson current is predominantly carried by Andreev bound states with high transmittance, and the fact that the skewness persists in junctions that are longer than the mean free path suggests that the effect may be related to the helical nature of the Andreev bound states in the surface of HgTe.**

Topological insulators (TI) have a special band structure with important consequences for proximity-induced superconductivity. In 3-dimentional topological insulators (3D-TI), the inversion of the conduction and valence bands leads to conducting 2D surface states with energies that are linearly proportional to their momenta [1-5]. Spin-momentum locking protects the charge carriers at the surface against elastic backscattering [6,7]. These special properties are reflected in the superconducting proximity effect in an S/3D-TI bilayer or an S/TI/S junction, which may host Majorana fermions in a quasi-1D channel or vortex core [8-10]. Most previous works characterized current-voltage characteristics to determine the critical current's dependence on temperature, gate voltage, or magnetic field [11-22], while a few studies characterized the CPR [23,24].

Here, we use a scanning SQUID microscope to perform contactless measurements of the diamagnetic response of Nb/HgTe bilayers and of the CPR of Nb/HgTe/Nb junctions. In contrast to previous CPR results [23,24], we find no evidence for bulk states, and we observe that the CPRs of many junctions of different sizes consistently exhibit forward skewness.

The CPR in an S/TI/S junction is a key diagnostic [8,25-32]. Weak disorder in the TI far from the superconducting contacts theoretically does not affect the induced superconducting state [33,34]; therefore, Andreev bound states should form in high-transmittance surface channels [8,26,27,29,31]. A CPR with forward skewness – that is, a deviation from a perfect sinusoidal form – is a signature of such high-transmittance Andreev bound states [35-37].

To our knowledge, there have not been direct observations of forward skewed CPRs in topological insulators [23,24], although the skewness has been indirectly inferred [24] from the Fraunhofer interference pattern. Previous CPR experiments in topological insulators [23,24] were complicated in part by bulk states, self-inductance effects, and bias voltage, factors that are eliminated in this work.

Moreover, a skewed CPR can also result from ballistic transport [35]. Measurements in metallic break junctions showed that the CPR approaches the predictions for quantum point contacts in the ballistic limit [38]. In metallic atomic point contacts, the CPR was significantly skewed only in contacts with very high transmittance channels [36]. There are reports of skewed current phase relations in graphene [39,40]. A skewed CPR was also observed in Josephson junctions based on a two-dimensional electron gas in semiconducting InAs with an electron mean free path comparable to the junction size [41]. Any observation of a skewed CPR in a S/TI/S junction will thus have to be scrutinized as to the origins of this particular shape.

Unstrained bulk HgTe is a semimetal that is charge-neutral when the Fermi energy is at the touching point between the light-hole and heavy-hole bands at the Brillouin zone center [42]. Epitaxial HgTe layers may readily be turned into a topological insulator by inducing strain in the material [42]. In contrast to Bi-compounds, such layers exhibit no bulk conductance [42,43]. Our samples are 65-nm-thick HgTe grown by molecular beam epitaxy under coherent strain on a CdTe substrate, leading to a 3D-TI with a full gap of ~22 meV.

We measured the CPR in Nb/3D-HgTe/Nb junctions (Fig. 1) using a scanning SQUID microscope. The field coil of the SQUID sensor threads magnetic flux, $\Phi_a$, through a superconducting ring with a Nb/3D-HgTe/Nb junction; this induces a phase drop, $\varphi/2\pi \approx \Phi_a/\Phi_0$, across the junction, where $\Phi_0$ is the superconducting flux quantum (see Supplementary information for the description of the calibration procedure). The SQUID sensor detects magnetic flux, $\Phi$, generated by the total current, $I$, in the ring. Both non-scanning [41,44,45] and scanning [23] variations of the present setup have been used in the past. The advantages of the scanning setup include the ability to measure a large number of samples in the same experiment, small parasitic inductances (which allow direct, low-noise measurements of CPR), and a natural magnetic background signal cancellation in gradiometric SQUIDs [23].

A typical measured non-sinusoidal CPR at a temperature of 400 mK is shown in Figure 1(c). In the following text, we refer to the deviation from the sinusoidal form as forward skewness [46]. More specifically, we define the skewness as the amplitude of the second harmonic in the Fourier decomposition of the CPR. Fitting the measured CPR to the functional form $I(\varphi) = I_C/\sqrt{1+A^2}(\sin(\varphi) - A\sin(2\varphi))$ determines both the skewness, $A$, and the critical current, $I_c$ (see Supplementary Information). Note that skewness may also be defined as the phase (modulo $2\pi$) at which the current is maximal; the values of the skewness estimated by these two methods are linearly proportional for small deviations of the CPRs from a perfect sinusoid.

The CPR depends on the junction width, $W$, defined as the superconducting line width (Fig. 2(a)). Theoretically, the critical current may increase with the number of conducting channels, given by $N = W k_F / \pi$, where $k_F$ is the Fermi wavevector. Figure 2(b) shows normalized CPRs, highlighting similarity in the functional form. Figure 2(c) shows that the critical current increases with the width of the junction, consistent with a larger number of conducting channels in wider junctions (see Supplementary Information). The normalized curves (Fig. 2(b)) and the dependence of the fitted skewness on width (Fig. 2(d)) both demonstrate that skewness is nearly identical for all $L$ = 200 nm samples, regardless of width.

We measured CPR in several junctions of length 200-600 nm (Fig. 3(a)), with junction length defined as the shortest distance between the superconducting leads. While the amplitude of the periodic CPRs decreased with junction length, the CPRs deviated from sine form even for the longest junctions shown in Figure 3(a). To emphasize this observation, we normalized the curves in Figure 3(a) by the critical current values (Fig. 3(b)). The critical current is plotted as a function of the junction length in Figure 3(c). The observed skewness varied from sample to sample, but persisted even in the longest junctions of 600 nm (Fig. 3(d)).

To test whether the observed proximity effect primarily occurs at the surface of the 3D-HgTe, we measured the magnetic susceptibility of Nb/HgTe and Nb/CdTe bilayers; mesoscopic Nb disks were fabricated on HgTe mesas or the bare, undoped CdTe substrate (Fig. 4(a)-(b)). In previous work, we found that the total magnetic susceptibility of superconducting Al layers fabricated on $Bi_2Se_3$ was much lower than that of similar Al layers fabricated on bare oxidized silicon [23], indicating that the inverse proximity effect of the bulk states in the $Bi_2Se_3$ suppressed superconductivity. The bulk of 3D-HgTe could potentially be conducting due to dislocations or doping; in the presence of high bulk conduction of the normal layer, the inverse proximity effect is expected to manifest as suppressed superfluid density of the bilayer [23,47-49]. However, in the absence of bulk conduction, the total superfluid density of a Nb/3D-HgTe bilayer is expected to be close to that of the Nb layer alone. In this case, there is no substantial suppression of the superfluid density due to the inverse proximity effect because the conducting surface layer is very thin compared to the Nb disk. According to this logic, the comparable susceptibility values of Nb/HgTe (Fig. 4 (c)) and Nb/CdTe bilayers (Fig. 4(d)), indicating no suppression of the total superfluid density due to the inverse proximity effect between Nb and strained HgTe, suggest that there is little or no bulk conduction in the HgTe.

In fact, the Nb/HgTe disks appear to have a slightly *higher* susceptibility compared to Nb/CdTe. This increased susceptibility could be due to the Nb/HgTe structure having a larger total superfluid density, either due to proximity effect or to surface morphology. However, it could also be due to the fact that the Nb/HgTe disks are slightly closer to the SQUID sensor, by ~100 nm. Supplementary Information contains the estimated susceptibility at different heights of the SQUID sensor; more detailed modeling and height calibration would be needed to distinguish between these possibilities. The absence of signatures of the possible inverse proximity effect in the superfluid density agrees with the previous observation of an integer quantum Hall effect in the normal state, showing that both the normal and the superconducting conductance of HgTe are dominated by the surface states rather than by bulk states [42].

S/normal metal/S junctions are characterized by the quasiparticle mean free path, $\ell$, and the superconducting coherence length, $\xi$ (both of the normal metal region). Theoretical work is simplified for junctions that are either short ($L << \xi$) or long ($L >> \xi$), and for junctions that are either ballistic ($L << \ell$) or diffusive ($L >> \ell$). The clean-limit coherence length is $\xi_0 = \hbar v_F / \pi \Delta$, where $\hbar$ is the Planck constant, $v_F$ is the Fermi velocity in the TI, and $\Delta$ is the induced gap. When $\xi_0 \geq \ell$, it is natural to define an effective coherence length, $\xi$, which becomes $\xi \sim (\xi_0 \ell)^{1/2}$ in the limit $\xi_0 >> \ell$ [35].

Based on parameters determined from transport measurements, junctions with $L$=200 nm are the best candidates to be in the limit of $\xi, \ell \gtrsim L$. Transport measurements on similar junctions indicate that the phenomenologically determined induced gap is $\Delta \approx 0.1 \Delta_{Nb} ... 0.2 \Delta_{Nb}$ [12,20], yielding a clean-limit superconducting coherence length $\xi_0 \approx$ 800 nm for $\Delta = 0.15$ meV. The quasi-particle mean free path is $\ell \approx 200$ nm as estimated from normal transport data (Supplementary Information).

To model the junctions theoretically, we consider S/3D-TI/S junctions, where the S region describes the surface state of the TI in the proximity with the s-wave superconductor; the 3D-TI is treated as a Dirac surface state with a single cone. As shown in [31], theoretically the Josephson current in such junctions is carried by helical Andreev bound states characterized by spin-momentum locking similar to the protected normal state. To determine whether our experimental results can be described within this framework, we extend the previous model [31] to junctions of finite length (but still technically in the short junction limit), and fit the measured CPRs of the $L = 200$ nm junctions. Our fitting procedure allowed for two free parameters: the induced gap $\Delta$ and the Fermi wavevector $k_F$, which together with the width determines the number of conducting channels, $N$. The transmission of the weak link was assumed to be one, based on a previous normal state data [20].

The data for 200 nm-long junctions is well described by the ballistic junction model with a Dirac surface state, as shown in Figure 2(a) (The full expression for the theoretical CPR is given in the Supplementary Information). Best-fit values of induced gap ranged from $\sim 0.12$ meV to $\sim 0.19$ meV, consistent with previously reported values in similar HgTe/Nb Josephson junctions [12,20]. The reduction of the fitted gap relative to the Nb gap ($\sim 1$ meV) has previously been shown [50] to be related to the mismatch of the Fermi surface parameters. The variability in the fitted gap may be primarily due to differences in the interface transparency. The best-fit number of channels for the L = 200 nm junctions agrees with the number of channels calculated using the geometrical width. The superconducting transport through these junctions is thus consistent with what is expected for a ballistic junction. Moreover, the involved surface states are known to be of a topological origin [20,42]. Both the observed skewness and the amplitude of the CPR agree with the model of ballistic quasiparticle transport through helical Andreev bound states [31]. Detailed descriptions of the induced gap and number of channels inferred from the fits are given in the Supplementary Information, together with the fit confidence range.

From the quasiparticle perspective, the forward skewness in the CPR means that highly transmitting Andreev bound states are important contributors to the Josephson current. Although the theory [31] that we used for our fits is not strictly applicable to all of the measured junctions, the persistent skewness in longer junctions indicates that highly transmitting Andreev bound states continue to dominate the transport even in junctions with $L = 600$ nm. This finding suggests that scattering in the induced superconducting state (in the TI region between the leads, rather than at the TI/S interface) is similar to that in the normal state transport and may be influenced by the helical nature of the Andreev bound states.

In summary, we have directly measured CPRs with forward skewness in Nb/3D-HgTe/Nb junctions. The skewness is present in the CPRs of all 16 measured junctions, indicating the importance of highly transmitting Andreev bound states for the superconducting properties of these junctions. Although other models may produce similar skewness, a theoretical model of S/3D-TI/S junctions fits well to the CPRs of the 200 nm-long junctions, suggesting that the helical nature of the Andreev bound states may be important for suppression of backscattering, similar to the spin-momentum locking in the normal state. Further expansion of the parameter space – faster measurement and smaller junction dimensions – will advance the study of unconventional nature of Andreev bound states, such as helical nature, in the proximity effect in S/3D-TI/S junctions by reducing the role of conventional modes and dynamic processes.

**Acknowledgements**

The work at Stanford University was supported by the Gordon and Betty Moore Foundation through grant GBMF3429. C. A. W., J. R. K. and K. A. M. acknowledge support from the Department of Energy, Office of Basic Energy Sciences, Division of Materials Sciences and Engineering, under Contract No. DE-AC02-76SF00515. E. M. H. acknowledges financial support from the German research foundation (DFG) within FOR1162 (HA5893/5-2). G.T. acknowledges financial support through the DGF grant No TK60/1-1. The work at EP3 was supported by the DARPA Microsystems Technology Office, MesoDynamic Architecture Program (MESO) through contract number N66001-11-1-4105, by the German Research Foundation (DFG Schwerpunkt 1666 'Topological Insulators', the DFG-JST joint research project 'Topological Electronics' and the Leibniz program) and the EU ERC-AG program (Project 3-TOP). The authors thank Roni Ilan, Fernando de Juan, Raquel Queiroz, Pavel Ostrovsky, and Martin Leijnse for fruitful discussions.

**Figure Captions**

Fig. 1. (Color online). Forward skewness in the CPRs of Nb/3D-HgTe/Nb junctions. (a) False-color scanning electron microscopy images of a representative Nb ring with a Nb/3D-HgTe/Nb junction. The separation between the superconducting edges, $L$, and width, $W$, are 200 nm and 1000 nm, respectively. (b) Schematic of CPR measurement. A current applied through the field coil produces an applied flux, $\Phi_a$, through the sample ring. The applied flux creates a phase drop, $\varphi$, across the junction, inducing a supercurrent, $I$, in the ring. The supercurrent modifies the magnetic flux through the pickup loop of the SQUID sensor (red). (c) The CPRs of this junction and three nominally identical ones (colored symbols) exhibit forward skewness, in contrast to a perfectly symmetric sinusoidal form (black solid line).

Fig. 2. Width dependence of the CPR. (a) CPR in Nb/3D-HgTe/Nb junctions with width $W = 300 - 1000 \, \text{nm}$ and $L = 200 \, \text{nm}$, with fits (solid lines) to the S/3D-TI/S model in the text. (b) Data from (a) normalized by the fitted critical current; the same functional form is evident independent of width. (c) Fitted critical current versus junction width for several samples. (d) Fitted skewness versus width. Vertical error bars in (c) and (d) are 68% confidence range of the fits of $I(\varphi) = I_C / \sqrt{1 + A^2} (\sin(\varphi) - A\sin(2\varphi))$ (see text); horizontal error bars are estimates of lithography edge imperfections observed in the SEM images (e.g. Fig. 1(a)). Skewness does not vary substantially with width.

Fig. 3. Length dependence of the CPR. (a) CPR in Nb/3D-HgTe/Nb junctions with nominal length $L = 200 - 600 \, \text{nm}$ and $W = 1000 \, \text{nm}$. (b) Data from (a) normalized by the fitted critical current. (c) Critical current versus junction length. Skewness (d) versus junction length (solid circles). Vertical error bars in (c) and (d) are 68% confidence range of fits of $I(\varphi) = I_C / \sqrt{1 + A^2} (\sin(\varphi) - A\sin(2\varphi))$ (see text); horizontal error bars are estimates of lithography edge imperfections observed in the SEM images (e.g. Fig. 1(a)). The skewness persisted even in the longest measured junctions.

Fig. 4. (Color online) Similar values of susceptibility were determined for HgTe/Nb and CdTe/Nb bilayers, showing no evidence of an inverse proximity effect in Nb on HgTe, and supporting the assertion that conducting states are absent in the bulk of the strained HgTe (see text). Comparing the magnetic susceptibility. (a), (b) False-color scanning electron microscopy images of typical Nb disks 2 μm in diameter on (a) a HgTe mesa and (b) CdTe. (c), (d) Susceptibility images and cross-sections of Nb on (c) HgTe and (d) CdTe. Each row in (c) and (d) represents a different sample. Orange dashed lines show the location of the cross-sections. Left-to-right skewness of the cross sections is due to the leads and alignment of the SQUID sensor. No evidence for the superfluid density suppression in HgTe/Nb bilayers is found.

Figure 1

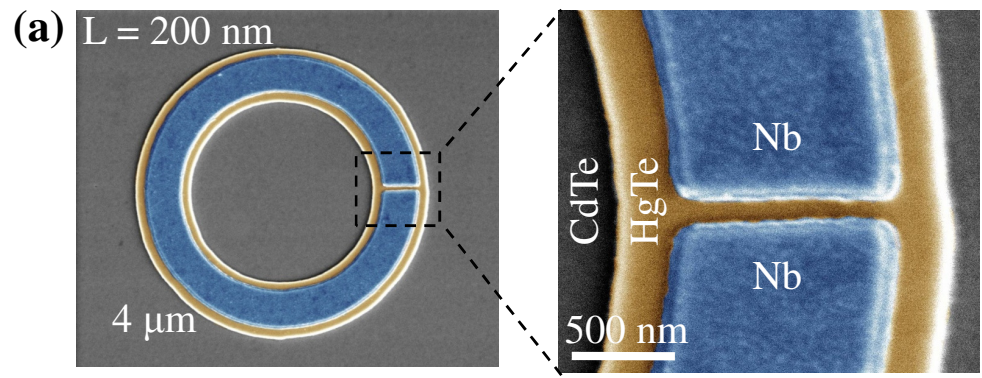

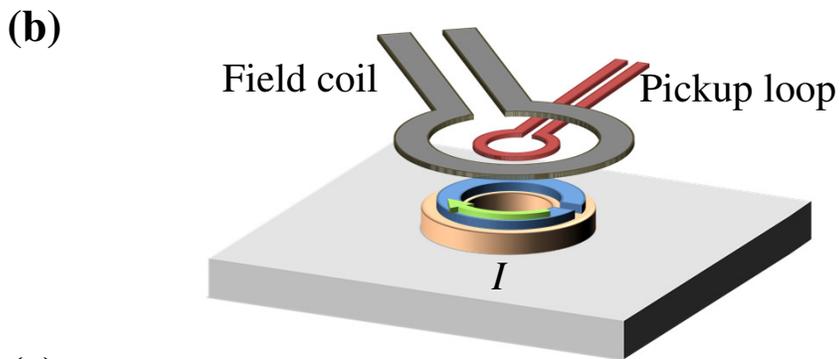

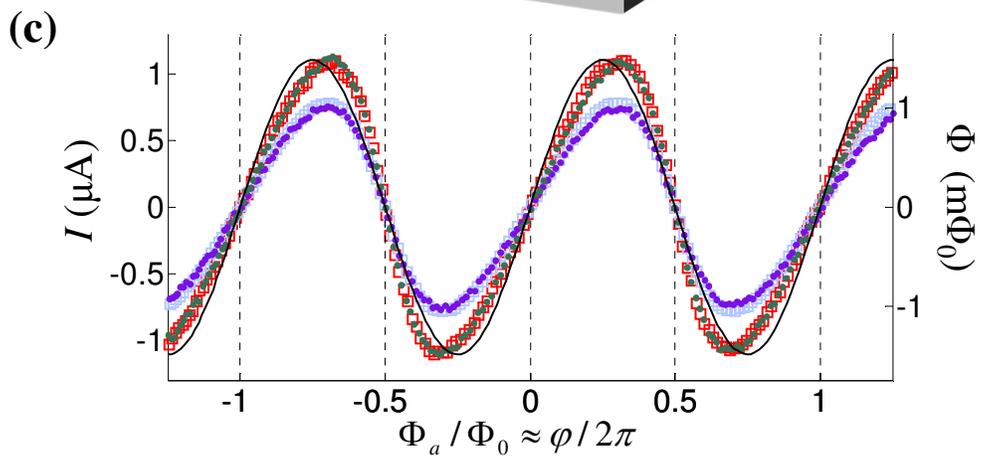

Figure 2

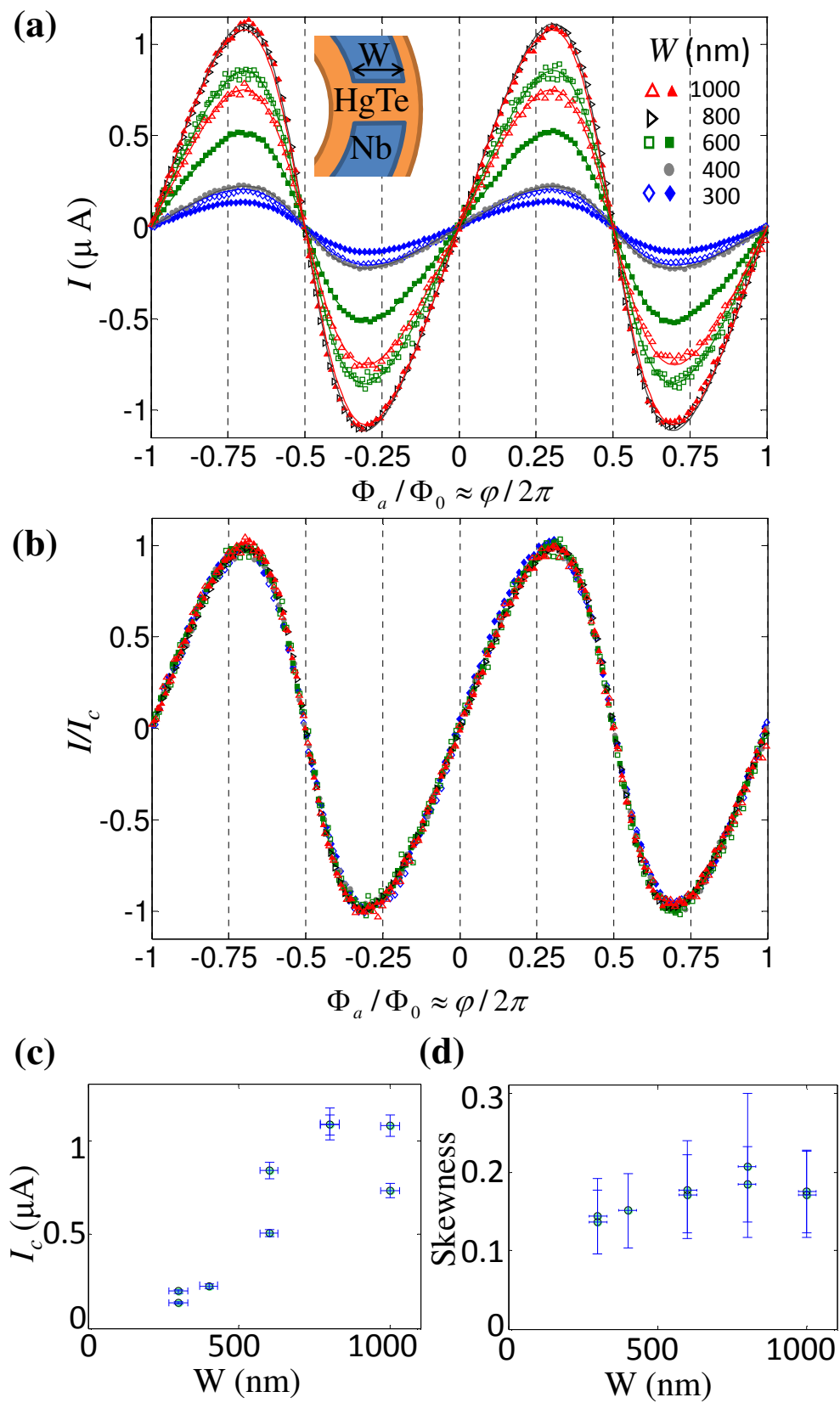

Figure 3

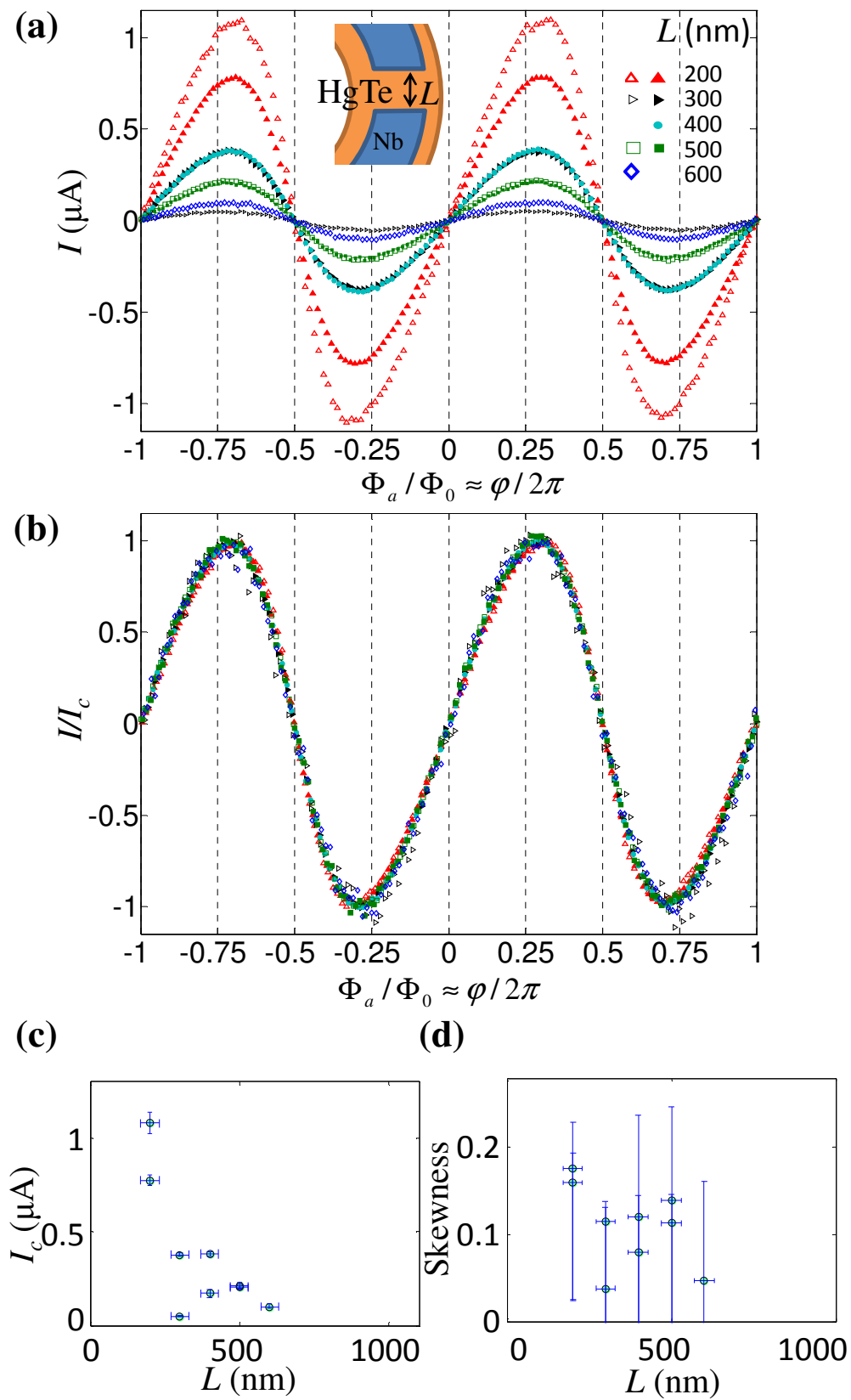

Figure 4

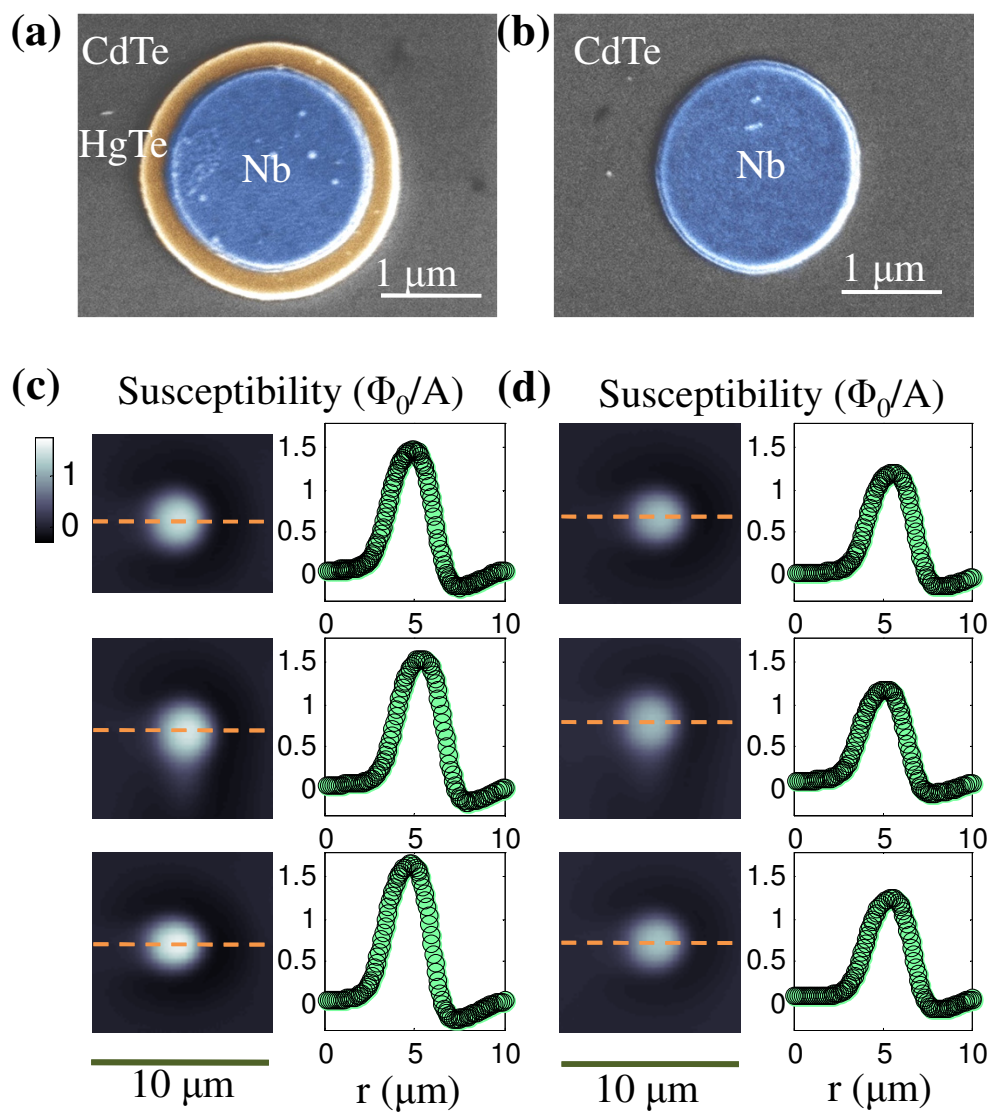

Supplementary Online Material for

# Non-sinusoidal current-phase relationship in Josephson junctions from the 3D topological insulator HgTe

Ilya Sochnikov[1,2], Luis Maier[4], Christopher A. Watson[1,3], John R. Kirtley[2,3], Charles Gould[4], Grigory Tkachov[5], Ewelina M. Hankiewicz[5], Christoph Brüne[4], Hartmut Buhmann[4], Laurens W. Molenkamp[4], and Kathryn A. Moler[1,2,3]

[1]*Department of Applied Physics, Stanford University, Stanford, California 94305, USA*

[2]*Geballe Laboratory for Advanced Materials, Stanford University, Stanford, California 94305, USA*

[3]*Stanford Institute for Materials and Energy Sciences, SLAC National Accelerator Laboratory, Menlo Park, California 94025, USA*

[4]*Physikalisches Institut (EP3), University of Würzburg, Am Hubland, 97074 Würzburg, Germany*

[5]*Institute for Theoretical Physics and Astrophysics, University of Würzburg, Am Hubland, 97074 Würzburg, Germany*

**Device fabrication**

This section describes the fabrication of Nb and HgTe hybrid structures. First, a 65 nm-thick HgTe layer was grown by molecular beam epitaxy on a CdTe substrate. The wafer was partially covered with a Ti etch mask then ion-milled with Ar to obtain mesa structures of HgTe. On the top surface of the HgTe mesas, 70 nm-thick Nb superconducting films were deposited in an ultra-high vacuum sputtering chamber followed by a lift-off process. Details of the fabrication steps appear in the caption to FIG. S1.

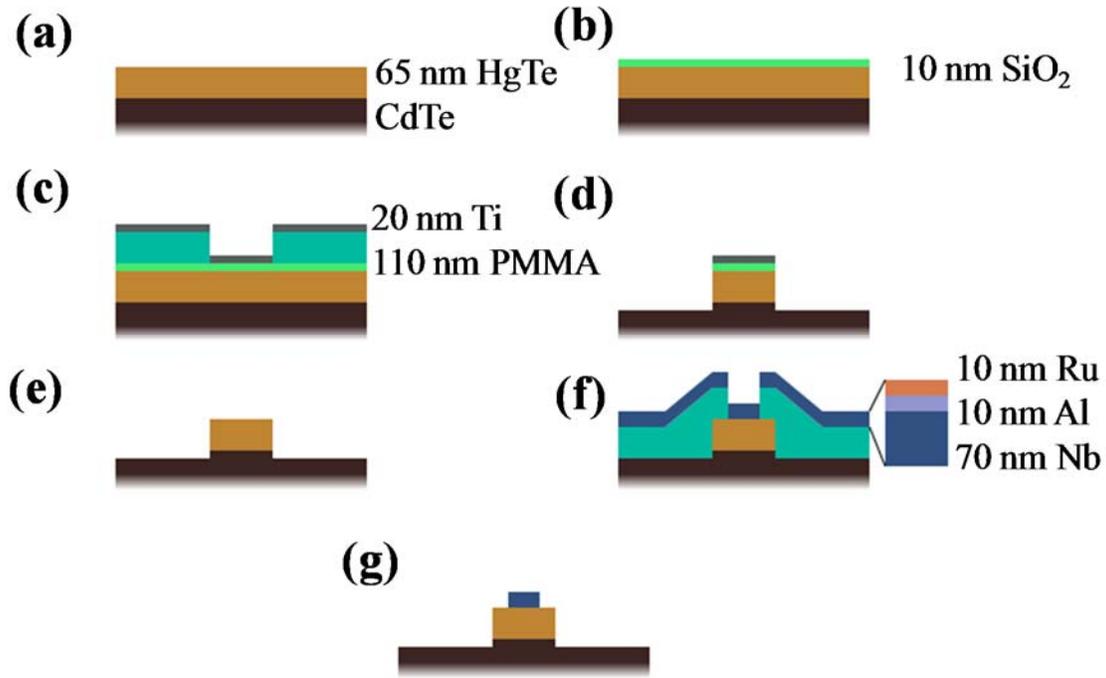

**FIG. S1. Fabrication of hybrid structures of superconducting Nb and the topological insulator 3D-HgTe.** (a) A 65 nm-thick HgTe layer is grown by molecular beam epitaxy on an insulating CdTe substrate. The lattice mismatch between HgTe and CdTe results in strain in the HgTe layer. (b) Ten nanometers of $SiO_2$ is grown using plasma-enhanced chemical vapor deposition. (c) The sample is spin coated with Poly(methyl methacrylate) (PMMA) 950 K 3%. The PMMA is baked, and exposed in a 2.5-keV e-beam writer to produce the mesa patterns. Twenty nanometers of Ti is deposited on the developed PMMA pattern. (d) The Ti layer is lifted off. $SiO_2$ in exposed areas is removed via reactive-ion etching with a mixture of $CHF_3$ and $O_2$. Afterward, the structure is cleaned with $O_2$ plasma. The areas of open HgTe are ion milled. (e) Ti and $SiO_2$ are removed in 1:10 $HF:H_2O$. The HgTe mesa is defined. (f) The sample is spin coated with PMMA 950 K 3%. The PMMA layer is exposed in a 2.5-keV e-beam writer, and developed in isopropanol. The HgTe interface is cleaned in a short Ar ion milling step, followed by deposition of 70 nm Nb, 10 nm Al, and 10 nm Ru. (g) Lift-off of the superconducting layer in acetone defines the final hybrid Nb/HgTe structure.

**Characterization of normal-state transport**

Characterization of normal-state electronic transport provides information on the carrier density, mobility, expected electron mean free path, and surface nature of conducting states in strained HgTe. To characterize normal-state electronic transport in the 65 nm-thick strained HgTe layer, we fabricated (from the same wafer as the superconducting devices described in the main text) a six-terminal Hall bar device with a channel length of 600 μm and width of 200 μm and measured the magnetic field dependence of its Hall resistance at $T = 4\,\text{K}$ (FIG. S2). From the low-field data, the mean mobility and combined charge density of the conduction electrons are extracted: $\mu \approx 47{,}000\,\text{cm}^2/(\text{V}\cdot\text{s})$ and $n = 6.6\cdot 10^{11}\,\text{cm}^{-2}$ (including both surfaces). This density leads to the approximate Fermi wavevector value of $k_F = 2\sqrt{\pi n_{2D}} \approx 0.2\,\text{nm}^{-1}$ (we divided the above carrier density by 2, assuming an equal density of carriers for the two surfaces) [1]. This value of the Fermi wavevector corresponds to an electron mean free path of $\ell = \mu_{TSS}\hbar k_F/e \approx 200\,\text{nm}$, where the mobility of the top surface state is $\mu_{TSS} \approx 1/3\,\mu$ [1]. Interestingly, the resistance traces in FIG. S2 already show quantization effects at these relatively high temperatures in magnetic fields above $B \approx 2\,\text{T}$. At lower temperatures, clear quantization behavior as in ref. [2] is expected. This observation provides supporting evidence that transport occurs effectively through 2D states at low temperatures, particularly below $T = 1\,\text{K}$.

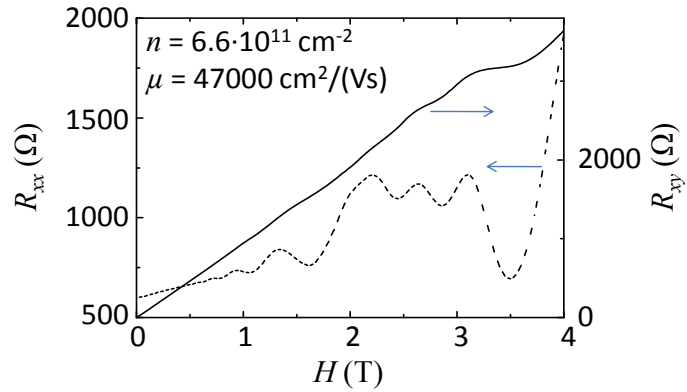

FIG. S2. **Magnetic field dependences of normal-state resistances.** Transverse, $R_{xy}$ (right ordinate as indicated by the arrow), and longitudinal, $R_{xx}$ (left ordinate as indicated by the arrow), resistances of a Hall bar fabricated from the same wafer as the Nb/HgTe structures described in the main text. The mobility, μ, and the carrier concentration, n, are extracted using the Drude equations [2].

**SQUID measurements**

*SQUID apparatus and operation mode*

The main component of our measurement setup is the SQUID susceptometer with two field coil and pickup loop pairs in a gradiometric configuration [3]; Figure 1(c) of the main text shows only one pickup loop and field coil. The scanning apparatus was mounted in a $^3$He cryostat, with a sample base temperature of ~450 mK, the temperature at which most of our measurements were performed. The white-noise floor of our scanning SQUID system is below 1 $\mu\Phi_0/\sqrt{Hz}$ [3].

In this work, we primarily operated the SQUID sensors in susceptometry mode [4,5]. The local magnetic field was swept relatively slowly by ramping the current through the field coil, inducing a local magnetic field [6]. We simultaneously measured the flux through the SQUID's pickup loop.

*Josephson current in a ring as a function of phase and flux*

To directly measure the current-phase relation (CPR), the junctions must be non-hysteretic, as specified in this section, and emphasis must be placed on determining the inductances of the SQUID's components and the sample rings. In our experiment, the magnetic flux produced by the supercurrent $I(\Phi)$ in a superconducting loop interrupted by a single Josephson junction is detected by the pickup loop. A proportionality factor between the detected flux and the value of the concentric supercurrent can be found using the procedure described below.

It is often difficult to prepare superconducting rings with small inductances and simultaneously have Josephson junctions with small critical current. Reaching these limits is important because for $2\pi L_S I_C / \Phi_0 \sim 1$, the junctions become hysteretic, leading to skewed current-flux curves. $L_S$ is the self-inductance of the measured ring, $\Phi_0 = h/2e$ is the superconducting flux quantum, $h$ is Planck's constant, and $e$ is the electron charge. These skewed current-flux curves could be confused with skewed current-phase curves of non-hysteretic junctions; thus, one should be careful when interpreting current-flux relations as CPRs, as the origin of asymmetry in the hysteretic case is not intrinsic to the Josephson current [7].

The calculations below show that the asymmetry in the measured CPRs is not caused by self-inductance effects, and that an approximation, $\varphi \approx 2\pi \Phi_a / \Phi_0$, can be made for our samples because they are characterized by small products of the inductance and the critical current, $L_S I_C$. This approximation allows direct measurements of the CPR.

In a superconducting ring with a Josephson junction, the amount of applied magnetic flux threading the ring, $\Phi_a$, will enforce a phase difference across the Josephson junction in the ring of $\varphi = 2\pi(\Phi_a + L_S I)/\Phi_0$, with $I$ being the induced supercurrent in the structure. This dependence is taken into account in the procedure below for assessing the CPR in a Josephson junction.

Consider a CPR given by

$$I(\varphi) = \frac{I_C}{\sqrt{1+A^2}}(\sin(\varphi) - A\sin(2\varphi)), \tag{S1}$$

where an $A\sin(2\varphi)$ term is introduced in addition to the conventional $\sin(\varphi)$ term in order to quantify the asymmetry observed in the measured current-flux curves.

Following Sigrist and Rice [7], the free energy of the junction is

$$U = \frac{\Phi_0}{2\pi}\int_0^\varphi d\varphi' I(\varphi') . \tag{S2}$$

The kinetic energy of the supercurrents in the ring is given by $L_S I^2/2$. The total free energy is therefore

$$F(I,\Phi_x) = \frac{L_S I^2}{2} + \frac{I_c \Phi_0}{2\pi\sqrt{1+A^2}}(1-\cos(\varphi)) + \frac{I_c \Phi_0 A}{2\pi\sqrt{1+A^2}}(1-\cos(\varphi)^2). \tag{S3}$$

Using $\varphi = 2\pi(\Phi_a + L_S I)/\Phi_0$ leads to

$$F(I,\Phi_x) = \frac{\Phi_0^2}{(2\pi)^2 L_S}\left[\frac{\phi_r^2}{2} + \frac{\beta}{\sqrt{1+A^2}}(1-\cos(\phi_x+\phi_r)) + \frac{\beta A}{\sqrt{1+A^2}}(1-\cos(\phi_x+\phi_r)^2)\right], \tag{S4}$$

where $\beta = \frac{2\pi L_S I_c}{\Phi_0}$, $\phi_x = \frac{2\pi\Phi_a}{\Phi_0}$, and $\phi_r = \frac{2\pi L_S I}{\Phi_0}$.

Using the ring self-inductance values as calculated in the next section, minimizing the free energy with respect to $\phi_r$ for various $\phi_x$ and using $I/I_c = \phi_r/\beta$ yields a current versus applied flux curve for given $\beta$ and $A$.

By fitting these curves using the estimated self-inductances of the rings (see below), we find the best-fit value of the only free parameter, $A$. This fitted value of $A$ versus $L$ is plotted in Figure 2(d), and versus $W$ in Figure 3(d). The fitted curves (Eq. S4) to the data from Figure 2(a) and Figure 3(a) appear in the top panels of FIG. S3(a) and S3(b), respectively. A small linear background is subtracted prior to the fitting procedure [8].

For comparison, the fits to the Sigrist-Rice model [7] in which the analog of Eq. (S4) is derived with a sinusoidal CPR are shown by the dashed lines in the lower panels of FIG. S3(a) and 3(b); these lines clearly indicate that the self-inductance effects are too small to explain the observed asymmetry, as mentioned in the beginning of this section.

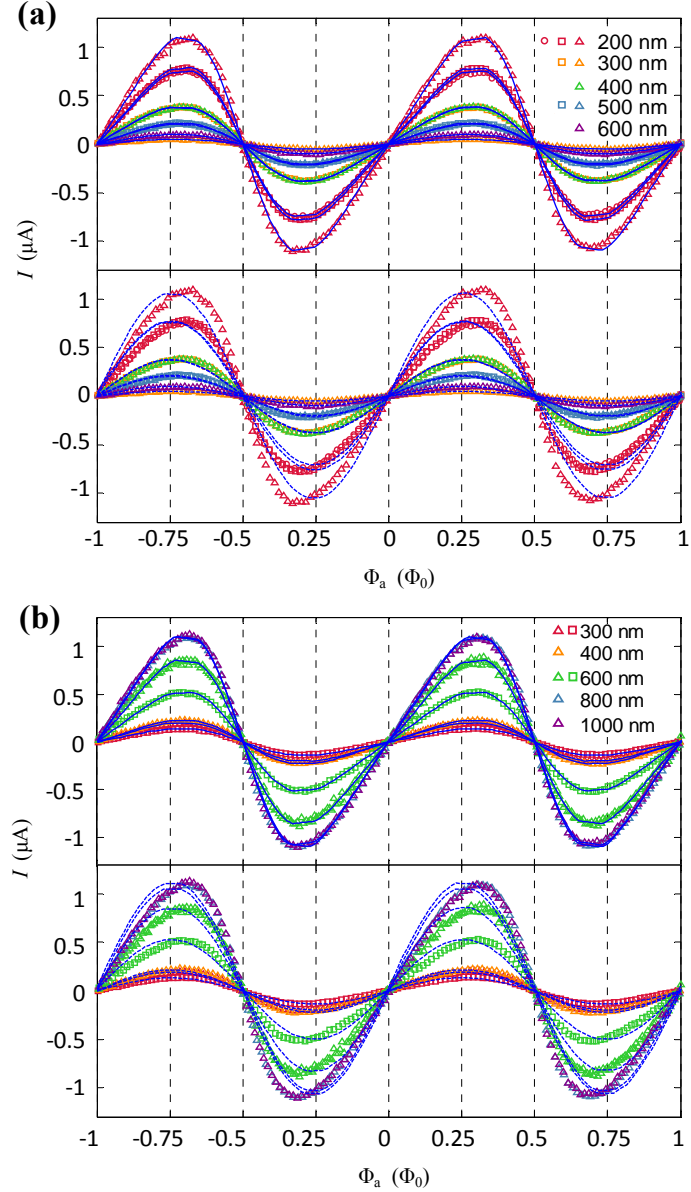

**FIG. S3. Fits of Eq. (S4) with non-sinusoidal CPR and of the Sigrist-Rice model to the measured current-flux curves.** (a) The upper and lower panels show fitted curves of Eq. (S4), and the Sigrist-Rice model, respectively, to data obtained for $L = 200$ nm rings with different junction widths. (b) The upper and lower panels show fitted curves of Eq. (S4) and the Sigrist-Rice model respectively, to data obtained for $W = 1000$ nm rings with different junction lengths. Clearly, differences in the fits qualities show that self-inductance effects alone cannot explain the skewness in the CPR.

*Calculations of inductances*

The current, $I$, flowing around the ring, which is proportional to the flux detected by the SQUID sensor pickup loop, $\Phi$, is given by

$$I = M_{S-PUL} \Phi, \quad (S5)$$

where $M_{S-PUL}$ is the mutual inductance between the sample ring and the pickup loop.

The current applied to the field coil produces flux in the ring:

$$I_{FC} = M_{S-FC} \Phi_x, \quad (S6)$$

where $M_{S-FC}$ is the mutual inductance between the field coil and the sample ring. The self-inductance, $L_S$, of the sample ring is needed to evaluate the flux generated by the circulating supercurrent in the sample ring: $\phi_r = \dfrac{2\pi L_S I}{\Phi_0}$, as mentioned in the previous section.

Inductances $M_{S-PUL}$, $L_S$, and $M_{S-FC}$, calculated using finite-element analysis, are shown in FIG. S4(a). The method is based on the work by Brandt [9] with the prescription for calculation of the Laplacian operator given by Bobenko and Springborn [10]. The pickup loop is modeled as a 1D wire with an effective radius $r_0 = 2.6$ μm. The sample ring is modeled as a thin-film superconductor with thickness $d$ smaller than the Pearl penetration depth, $\Lambda$. The model is valid for an arbitrary value of the Pearl penetration depth relative to the sample lateral dimensions (ring radius or superconducting strip width), which may vary in our samples. The Pearl penetration depth is defined as $\Lambda = 2\lambda^2/d$, where $\lambda$ is the London penetration depth of Nb. The London penetration depth of $\lambda = 150$ nm is estimated by comparing the results of the same model with the susceptibility of one of the rings as shown in Figure S4(b)-(d).

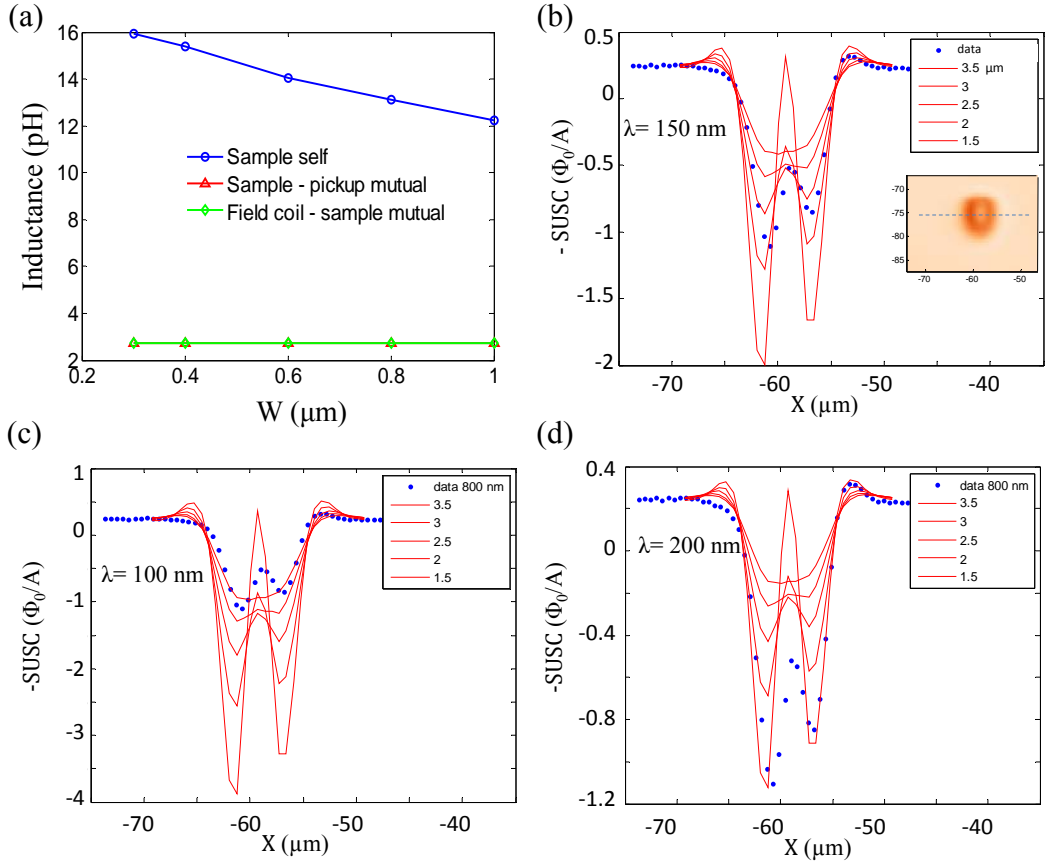

**FIG. S4**. **Sample self-inductances $L_S$, mutual inductances of the sample ring to the SQUID pickup loop $M_{S-PUL}$, and mutual inductances of the sample ring to the SQUID field coil $M_{S-FC}$, for various junction widths (assumed to be equal to the width of the superconducting Nb line).** (a) The curves shown are calculated for $\lambda = 150$ nm and SQUID sample separation of $\sim 2$ μm. The sample-pickup mutual inductance and field coil-sample mutual inductances do not depend on the ring's annulus width and the self-inductance of the sample changes slightly with the width. (b) Susceptibility of a rings with 800 nm annulus width shown as dots (this is a cross section along the dashed line in the inset image, scales in the inset are in micrometers) is compared to the calculated susceptibility. The solid lines are calculated for the sample to SQUID vertical separation of 3.5, 3, 2.5, 2, 1.5 μm, corresponding to lowest to highest amplitude in the solid lines. Penetration depth used is $\lambda = 150$ nm. (c)-(d) Same analysis as (b) but with $\lambda = 100$ nm and $\lambda = 200$ nm. A combination of sample to SQUID separation of $\sim 2$ μm, and $\lambda \approx 150$ nm can describe the observed susceptibility, yielding the values of inductances that are used for calibrations.

*Susceptibility of superconducting disks*

We simulate the susceptibility of the superconducting disks (experimental data in Fig. 4 of the main text) using the same finite-element analysis as in the preceding section. As mentioned above, the algorithm is based on the work by Brandt [9] with the prescription for calculation of the Laplacian operator given by Bobenko and Springborn [10].

The pickup loop is modeled as a 1D wire with an average radius $r_{mean}$ = 1.66 μm. This value is different from the effective radius defined by the sensitive area of the pickup loop partially including the leads; for structures smaller than the dimensions of the pickup loop, the mean radius reproduces the experimental data more accurately (data not shown). The results of the calculations are shown in FIG. S5.

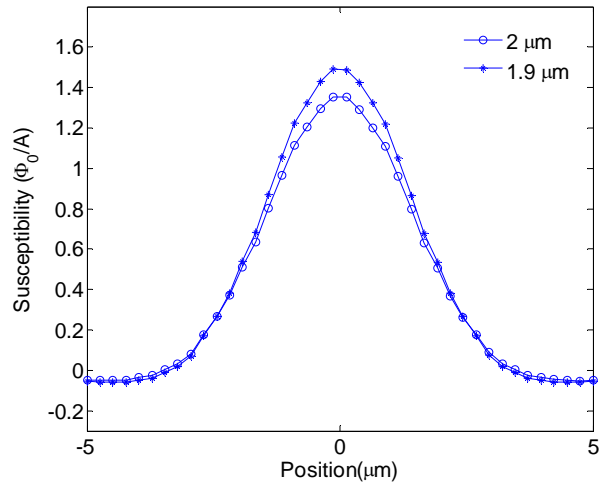

**FIG. S5. Simulated cross-sections of the susceptibility of 70 nm-thick superconducting disks with London penetration depth $\lambda$ = 150 nm for 2 μm and 1.9 μm separation between the sample and the SQUID pickup loop.** The magnitude of the susceptibility is consistent with the experimental data in Figure 4 of the main text. The difference of 100 nm in the separation of the sample from the pickup loop approximately reproduces the difference in the susceptibility observed between disks of Nb on Cd and disks of Nb on a ~100 nm HgTe mesa, supporting our claim of the absence of inverse proximity effect between Nb and HgTe.

These simulations show that the difference between the susceptibility of the Nb disks on CdTe and HgTe is explained by the height difference between the two types of structures. Thus, the susceptibilities of the Nb/CdTe and Nb/HgTe structures are approximately the same, with no evidence of suppression of the superfluid density in Nb/HgTe bilayers as would be expected from a possible inverse proximity effect.

**Dayem bridge model**

As mentioned earlier, inductance effects may distort a sinusoidal CPR in a Josephson junction. As shown above, the ring self-inductance is very small in our samples and does not contribute to the asymmetry in the CPRs in our experiments.

Kinetic inductance effects have been shown to cause an asymmetry in the CPR of a weak link [11,12]. We fit a weak-link (Dayem bridge) model [11] based on the Ginzburg-Landau equations to our CPR data (FIG. S6(a)). The extracted values of the free parameter $L_{fit}/\xi$, the ratio of the best-fit junction length to the coherence length, are shown in FIG. S6(b). The choice of this ratio does not impact the results. Assuming that $\xi$ does not change with length, the fitted length does not increase as the nominal length of the junction increases. Thus, no significant phase change occurs in our junctions due to kinetic induction of the weak link, suggesting that the asymmetry in our junctions is caused by the Andreev bound states with high transmittance rather than by inductance effects.

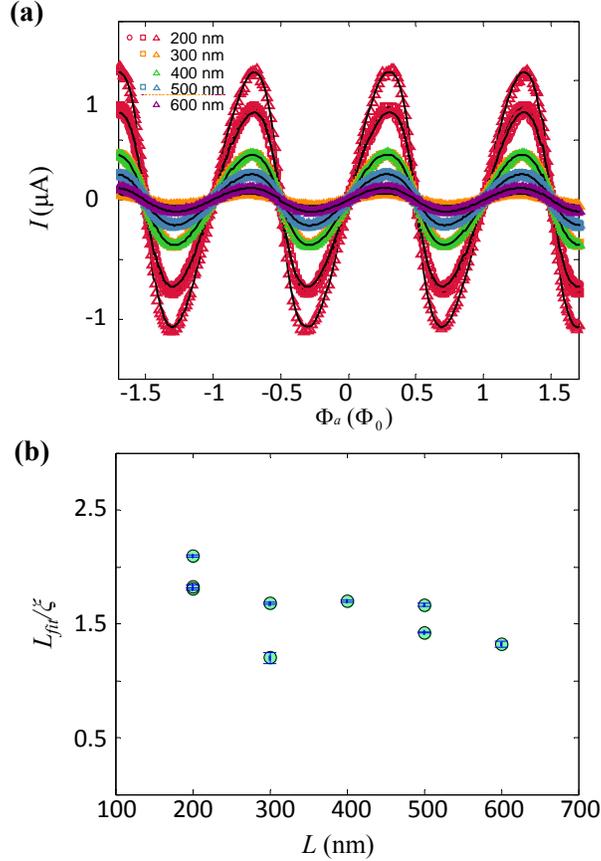

**FIG. S6. Dayem bridge model for a weak superconducting link.** (a) Full fits of the model of CPR in a superconducting weak link to the measured current-flux relations in junctions with varying length and the width 1000 nm . (b) Fitting parameter (ratio of the junction length to the coherence length) of the model plotted against the nominal length of the junctions. The fitting-parameter behavior disagrees with the expected increase for nominally longer junctions, showing that the observed CPR cannot be explained with a Dayem bridge model.

## CPR modeling for a ballistic superconductor S/3D-TI/S junction

*Model*

This section provides further details on our theoretical approach to interpreting the experimental results presented in the main text. We consider a Josephson junction created by depositing two conventional s-wave superconductors (S) on top of a 3D TI at a distance $L$ from each other. The junction can be described by the Bogoliubov-de Gennes Hamiltonian of the form

$$H_{BdG} = \begin{bmatrix} h(r) & i\sigma_y \Delta(x) e^{i\varphi(x)} \\ -i\sigma_y \Delta(x) e^{-i\varphi(x)} & -h^*(r) \end{bmatrix}, \quad h(r) = -i\hbar v(\sigma_x \partial_x + \sigma_y \partial_y) + U(x) - \mu. \quad (S7)$$

In the equations above, $h(r)$ is the surface Hamiltonian of the TI, where $\sigma_x$ and $\sigma_y$ denote spin Pauli matrices, $v$ and $\mu$ are the surface Fermi velocity and energy, respectively, and the potential $U(x)$ accounts for the likelihood of quasi-particle scattering. The off-diagonal entries in $H_{BdG}$ (S7) incorporate a spin-singlet s-wave pair potential induced in the TI underneath the S contacts. We assume that the superconducting gap $\Delta(x)$ and phase $\varphi(x)$ both vary across the Josephson junction according to

$$\Delta(x) = \begin{cases} 0, & |x| < L/2 \\ \Delta, & |x| \geq L/2 \end{cases}, \quad \varphi(x) = \begin{cases} -\varphi/2, & |x| \leq L/2 \\ +\varphi/2, & |x| \geq L/2 \end{cases} \quad (S8)$$

where $\varphi$ denotes the Josephson phase difference between the S terminals. In our Josephson junctions, $\varphi$ is controlled by a magnetic flux $\Phi_a$ enclosed in the superconducting Nb ring: $\varphi \approx 2\pi \Phi_a / \Phi_0$.

To calculate the Josephson current $I$ at finite temperature $T$, we adopt the approach in refs. [13,14], where $I$ is expressed in terms of a sum over the fermionic Matsubara frequencies $\omega_n = (2n+1)\pi k_B T$ (with $n = 0,1,\ldots$) as

$$I = -\frac{2e}{\hbar} k_B T \frac{\partial}{\partial \varphi} \sum_{\omega_n} \sum_{k_y} \ln D(\varepsilon, k_y, \varphi) \Big|_{\varepsilon = i\omega_n}. \quad (S9)$$

Another sum runs over all conducting channels with different transverse wavenumbers $k_y$. The energy levels of the system correspond to the zeros of the function $D(\varepsilon, k_y, \varphi)$. The latter is defined as the determinant of the eigenvalue equations obtained by matching the scattering states of $H_{BdG}$ at the boundaries $x = \pm L/2$ and the scattering region [13,14]. For Dirac surface states with $\mu \gg \Delta$ and $U(x) = U\delta(x)$, we find

$$D(\varepsilon, k_y, \varphi) = [1 - \alpha^2(\varepsilon) e^{i\beta(\varepsilon,k_y)}]^2 + 4\alpha^2(\varepsilon) e^{i\beta(\varepsilon,k_y)} t(k_y) \sin^2(\varphi/2). \quad (S10)$$

Here, $\alpha(\varepsilon)$ is the amplitude of the Andreev reflection at the S terminals:

$$\alpha(\varepsilon) = \frac{\Delta}{\varepsilon + i\sqrt{\Delta^2 - \varepsilon^2}}, \tag{S11}$$

$\beta(\varepsilon, k_y)$ is the phase difference between the particle and the hole after traversing the normal region:

$$\beta(\varepsilon, k_y) = \frac{2\varepsilon L}{\hbar v \cos\theta(k_y)}, \qquad \cos\theta(k_y) = \sqrt{1 - \frac{k_y^2}{k_F^2}}, \tag{S12}$$

where $\theta(k_y)$ is the angle between the particle momentum and the current direction, and $t(k_y)$ is the transmission probability through the normal region:

$$t(k_y) = \frac{\cos^2\theta(k_y)}{1 - \sin^2\theta(k_y)/(1+Z^2)}, \quad Z = \frac{U}{\hbar v}, \tag{S13}$$

with $Z$ being the dimensionless barrier strength.

The above equations yield the Josephson current as a function of the magnetic flux $\Phi_a$ and the temperature $T$ for a given junction length $L$, width $W$, induced superconducting gap $\Delta$, and the number of conducting channels $N$:

$$I(\Phi_{apl}, T) =$$

$$-\frac{4e}{\hbar} k_B T \sum_{n=0}^{\infty} \sum_{m=-N/2}^{N/2} \frac{\alpha^2(\varepsilon) e^{i\beta(\varepsilon,k_y)} t(k_y) \sin(2\pi\Phi_{apl}/\Phi_0)}{[1 - \alpha^2(\varepsilon) e^{i\beta(\varepsilon,k_y)}]^2 + 4\alpha^2(\varepsilon) e^{i\beta(\varepsilon,k_y)} t(k_y) \sin^2(\pi\Phi_{apl}/\Phi_0)} \Bigg|_{\substack{\varepsilon = i\omega_n \\ k_y = \frac{2\pi m}{W}}},$$

$$\tag{S14}$$

where we adopt periodic boundary condition in the $y$ direction, with $m$ being the channel index. As shown in the next sections, fits to the experimental data can be obtained by adjusting only two parameters, the induced gap $\Delta$ and the effective number of channels $N$. The band structure constant $\hbar v$ was set in the following sections to 318 meV·nm in accord with the $\mathbf{k} \cdot \mathbf{p}$ model for HgTe [2].

*Fits to S/3D-TI/S model*

This subsection describes the fitting procedure of the above S/3D-TI/S model to the measured CPRs. The model is, strictly speaking, applicable only to the case of short, clean junctions, with the length and the coherence length smaller than the electron mean free path, $(L, \xi_0) \ll \ell$. In accordance with our normal-state resistance measurements predicting an estimated mean free path of about two hundred nanometers and the coherence length $\xi_0 \approx 800$ nm (see the main text), we consider the junctions with $L = 200$ nm to be close to the appropriate limit for fitting the model. Therefore, we begin with the fits for the shortest measured junctions, $L = 200$ nm, with $W$ values varying between 300 nm and 1000 nm.

The model predicts a contribution from individual modes that are identified by a specific $k_y(m)$. The number of modes contributing to the supercurrent, when one of the TI surfaces carries a current, is $N = W k_F / \pi$. We emphasize that $k_F$ is sensitive to two assumptions; for the fits below, we assumed that $W$ is equal to the nominal width and that only one surface of HgTe carries the supercurrent. However, it is possible that both surfaces carry the supercurrent (not due to hybridization effects, but rather the simple possibility that pair correlations can be induced much more deeply under the Nb contact than just the top surface) in our samples; in general, fits with two surfaces would yield smaller $k_F$ values. The overall reduction of the superconducting gap underneath the Nb electrodes (due to finite transparency of the Nb/HgTe interface and possible disorder effects) is absorbed into the fit of the effective gap value [15,16].

To establish the most-likely values of fitting parameters for a given junction geometry, Equation (S14) is evaluated for a wide, coarse mesh of parameter space defined over the effective gap $\Delta$ and $k_F$. The number of modes is enforced by $N = W k_F / \pi$, with the additional constraint of fixing the total number to the nearest odd integer ($N$ even, with a mode at $k_y = 0$). A fine mesh is then defined, centered on the best fit values over the coarse space, for a final resolution of $0.002$ meV and $0.005$ nm$^{-1}$ for $\Delta$ and $k_F$, respectively, and large enough to enclose the 68% confidence interval, as described in the following. The root-mean-square residual defined at each point of the new mesh is used to generate a mean square residual ($\Xi^2$) contour map, with the $\Xi^2 = 2\Xi^2_{min}$ and $\Xi^2 = 5\Xi^2_{min}$ contours interpreted as the 68% and 95% confidence intervals, respectively. The errors and error bars reported for fitting parameters throughout this section correspond to the furthest extent of the 68% confidence interval. The confidence intervals are in general not simply connected, owing to local minima where $k_F$ is naturally consistent with an odd integer number of modes. In such cases, the errors enclose all such local minima that include $2\Xi^2_{min}$ contours. An example of the fitting results for a junction of $L = 200$ nm and $W = 1000$ nm appears in FIG. S7. All of the best fits for the $L = 200$ nm junctions are shown in Figure 2(a) of the main text.

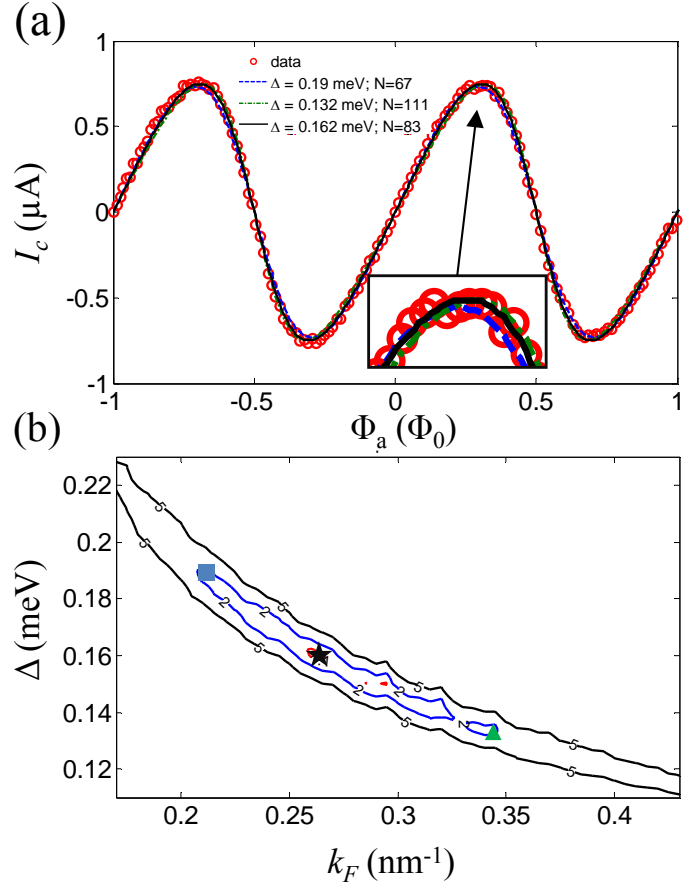

**FIG. S7. Determination of error bars for the fits of the S/3D-TI/S model.** (a) The best fit and the curves at the 68% confidence plotted for a junction of $L = 200$ nm and $W = 1000$ nm. (b) $\Xi^2/\Xi^2_{min}$ contours for $\Xi^2 = 1.1\Xi^2_{min}$ (red), $\Xi^2 = 2\Xi^2_{min}$ (blue), and $\Xi^2 = 5\Xi^2_{min}$ (black). The blue square, black star, and green triangle correspond to the blue solid, black dashed, and green dash-dot lines in (a). Small visual differences between the fits and the data (see inset in (a)) correspond to substantial variations in fitting parameters and hence a rather wide confidence region in (b).

FIG. S8 shows the best fit values and error bars for the extracted Fermi wavevector and the induced gap for the width series.

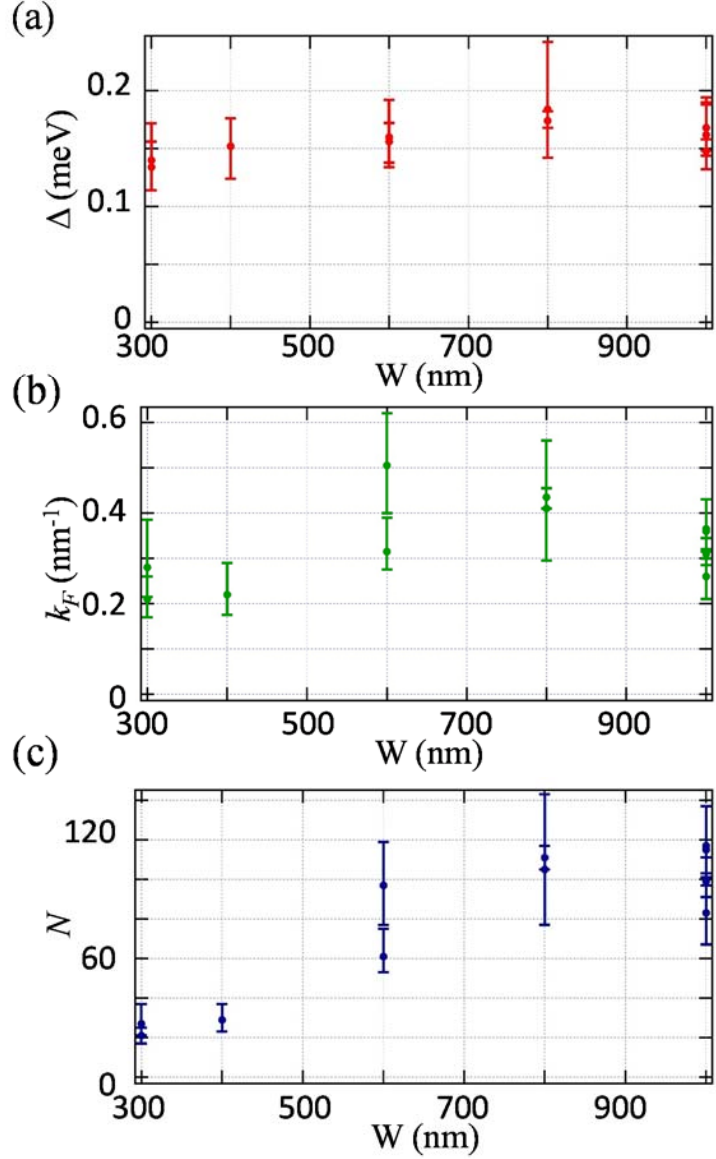

**FIG. S8. Values of the induced gap (a), Fermi wavevector (b), and the number of modes (c) producing the best fits of the S/3D-TI/S model to the CPRs measured in junctions with $L = 200$ nm and varying width.** The error bars represent the 68% confidence region of the fits. The model reproduces the skewness in the measured CPRs, and the number of channels increases with the width of the junctions.

In the width series, as mentioned earlier, the $k_F$ values are subject to a large uncertainty in the definition of the junction width and whether two or one surfaces of the TI carry the supercurrent. For these reasons, the extracted values cannot be compared conclusively with previously reported values [2]. However, in general, the values extracted from our fits do not differ greatly from the expected $k_F \approx 0.2$ nm$^{-1}$ from the normal transport data. Including two surfaces in the fits would yield smaller values of $k_F$.

Fitting the model to the longer junctions is a special case that requires an extended explanation. In the experiment, the critical current value drops quite significantly with the length. This decrease is due to the junctions passing beyond the strict short-junction limit, possibly in terms of both the coherence length and the mean free path. As more defects occur in the longer junctions, some channels become blocked or less transparent, leading to an overall drop in the critical current value. Nonetheless, the model is capable of capturing the asymmetry and the induced gap values in longer junctions, thus supporting our conclusion that the observed asymmetry is due to the formation of the Andreev bound states with high transmittance even in longer junctions.

FIG. S9 shows the best fit values and error bars for the extracted Fermi vector and the induced gap for the length series.

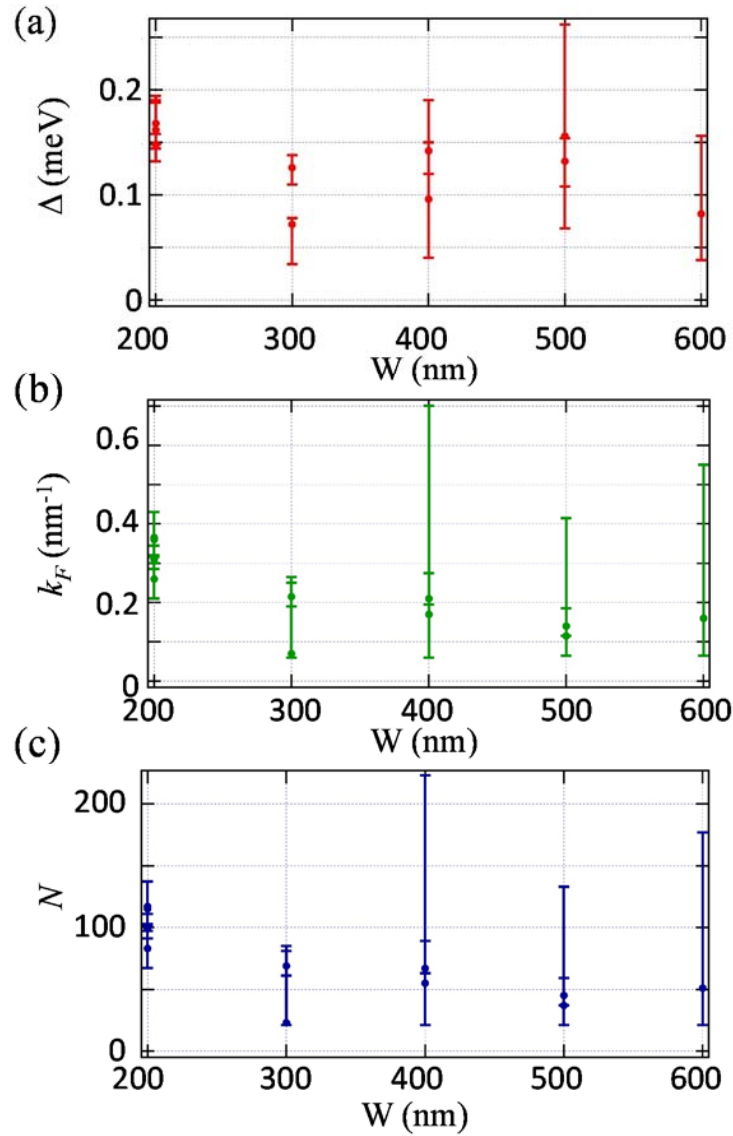

**FIG. S9. Values of the induced gap (a), Fermi wavevector (b), and the number of modes (c) producing the best fits of the S/3D-TI/S model to the CPRs measured in junctions of varying length and $W = 1000$ nm.** The error bars represent 68% confidence

region of the fits. The model reproduces the skewness in the longest junctions, although the number of channels is reduced with increasing length.

Although skewness is not unique to the model described here, our observation that this theoretical model of ballistic S/3D-TI/S junctions fits well to the CPRs of junctions that are 200 nm long suggests that the helical nature of the conducting channels may be important for suppressing backscattering in a way that is similar to that achieved in the normal state. Our observation that skewness persists to longer lengths indicates that highly transmitting Andreev bound states continue to dominate transport at all measured lengths.